\documentclass[12pt]{article}
\usepackage{amssymb}
\usepackage{graphicx}

\begin{document}
\thispagestyle{empty}
\begin{center}{\Large{\textbf{Dark photons and tachyonic instability induced by Barbero-Immirzi parameter and 
axion-torsion transmutation}}}
\end{center}
\vspace{0.0000000000000000000000000000000000000000000000000000000000000000000000000000000000000001cm}
\begin{center}
{\large Zhi-Fu Gao\footnote{Xinjiang Astronomical Observatory, Chinese Academy of Sciences, Urumqi, Xinjiang, 830011, 
China, zhifugao@xao.ac.cn;}}  
\end{center}

\begin{center}
{\large Biaopeng Li\footnotemark[1]\footnote{University of Chinese Academy of Sciences, Beijing, China;}}
\end{center}

\begin{center}
{\large L.C. Garcia de Andrade\footnote{Cosmology and gravitation group, Departamento de
F\'{\i}sica Te\'{o}rica - IF - UERJ - Rua S\~{a}o Francisco Xavier
524, Rio de Janeiro, RJ, Maracan\~{a}, CEP:20550. Institute for Cosmology and philosophy of Nature, Croatia. e-mail:luizandra795@gmail.com}}
\end{center}
\vspace{0.000000000000000000000000000000000000000000000000000000000000000000000000000000000000000000001cm}
\begin{abstract}
In this paper, we investigate Holst gravity by examining two distinct examples. The first example involves minimal coupling 
to torsion, while the second explores non-minimal coupling. The motivation for the first example stems from the recent work 
by Dombriz et al. (Phys. Lett. B, 834:137488, 2022), which utilized a technique of imposing constraint constant coefficients 
to massive torsion in the model Lagrangian to determine parameters for the Einstein-Cartan-Holst gravity. We extend this 
methodology to investigate dark photons, where axial torsion transforms into axions.Interest in elucidating the abundance of 
dark photons within the framework of general relativity was sparked by Agrawal et al. (Phys. Lett. B, 801:135136, 2020). 
Building on the work of Barman et al.( Phys. Rev. D. 101: 075017, 2020), who explored minimal coupling of massive torsion 
mediated by dark matter (DM) with light torsion on the order of 1.7 TeV, we have derived a Barbero-Immirzi (BI) parameter of 
approximately 0.775. This value falls within the range established by Panza et al. at TeV scales, specifically $0\le{\beta}
\le{1.185}$ (Phys. Rev. D. 90:125007,2014). This seems to our knowledge the first time BI parameter is induced by dark 
photons on a minimal EC gravity. Very recently, implications of findings of BI parameter in cosmological bounces has 
appeared in the literature (Phys. Dark. Univ. 44:141078, 2024). For a smaller BI parameter a higher torsion mass of 
$1.51$\,TeV is obtained. Nevertheless. this figure is still a signature of light torsion which can be compatible with light 
dark photon masses. Magnetic helicity instability of dark photons is investigated. Axion oscillation frequency is shown to 
depend on the BI parameter and the BI spectra is determined by an histogram. This study not only broadens the understanding of 
Holst gravity but also provides crucial insights into the interplay between torsion, dark photons, and axions in the 
cosmological context.
\end{abstract}
\newpage
\section{Introduction}
Previously Barman et al.\,\cite{1}, following results by Belayev et al.\,\cite{2} and Almeida et al.\,\cite{3}, 
investigated the mediation of dark fermionic matter by torsion. They used data from hadron physics colliders in 
the channels of quarks and anti-quarks. More recently, Rufrano and Lambiase have shown that when the Barbero-Immirzi (BI) 
parameter\,\cite{4} comes into play in the realm of Holst gravity\footnote{Holst gravity is an extension of general 
relativity proposed by Peter Holst in 1971. It modifies the Palatini action by adding a topological term known as the 
Nieh-Yan term, resulting in a new dynamically equivalent form. It is dynamically equivalent to standard general 
relativity in the absence of torsion. Holst gravity is significant in quantum gravity theories, particularly in Loop 
Quantum Gravity (LQG). The Hamiltonian formulation of Holst gravity is related to Ashtekar variables, which reformulate 
general relativity as a special type of Yang-Mills gauge theory.}, it could potentially solve some physical issues related 
to matter and antimatter. Previous cosmological bounce solutions of the Holst gravity term in the Einstein-Cartan (EC) action 
with contact spin-spin four-fermion interactions have been considered, leading up to more recent bouncing solutions in 
fermion condensates by Tukhanashvilli and Steinhardt\,\cite{5}. The BI parameter was previously found by Panza et al.\,\cite{6} at TeV scale hadron colliders such as Large Hadron Collider\,(LHC). Garcia de Andrade has very recently 
discovered dark torsion oscillations in bouncing cosmological settings\,\cite{7,8}.

In this paper, following the hadron collider data as in Barman et al.\,\cite{1} and the findings of the BI parameter in 
several frameworks—such as matter-antimatter asymmetry\,\cite{4} and as a dark torsion candidate in Einstein-Cartan-Holst\,
(ECH) gravity\,\cite{9} by Dombriz et al.\,\cite{10}, we show that the BI parameter can be constrained to small estimates 
of $0.803$, which is shown to be compatible with the light torsion of $1.7\,$TeV found by Barman et al.\,\cite{1}. To 
obtain this estimate, we utilize Shapiro’s\,\cite{11} minimal coupling between fermions and torsion, which can be expressed 
in terms of Dirac matrices by the coupling rule
\begin{equation}
{\partial}_{k}\rightarrow{[{\nabla}_{k}+i{\eta}_{1}{\gamma}^{5}S_{k}+i{\eta}_{2}T_{k}]} .
\label{1}
\end{equation}
In this new operator, $\nabla_{k}$ represents the covariant Riemannian derivative, ${\eta}_{A}$ (where $(A=1,2)$) represents 
the coupling of the axial pseudo-vector ${S}_{i}= {\epsilon}_{ijkl}T^{jkl}$ ($i,j,k=0,1,2,3$) and $T_{ijk}$ represents the 
components of the Cartan torsion tensor. The other coupling is associated with the trace of torsion given by $T_{k}=
T^{j}_{kj}$, and ${\gamma}^{5}$ is the gamma-5 matrix in Dirac algebra. 

A fundamental ingredient in this paper is the investigation of dark photons, recently studied in terms of their abundance 
in the universe by Agrawal et al.\,\cite{12}. Here, we show that their presence helps us to determine the coefficients in 
the Lagrangian density of the model. Since we are using Holst gravity, we have both terms of the torsion vector and 
pseudo-vector. By applying the operator\,(\ref{1}) to the axial torsion $S_{k}$, one gets
\begin{equation}
{\partial}_{k}S^{k}\rightarrow{[{\nabla}_{k}S^{k}+i{\eta}_{1}{\gamma}^{5}S^{2}+i{\eta}_{2}T_{k}S^{k}]} .
\label{2}
\end{equation}
Note that the third term on the right of the above equation, the Holst parity-violating term, appears naturally. Thus, 
it is clear that this operator helps us significantly in obtaining the estimates for the BI parameter induced by the 
dark photons. It is interesting to note that by Cartan equation the $S^{2}$ term is equivalent to the four-fermion 
interaction term $J^{2}_{5}$ of four-chiral fermion. This allows us to determine the BI parameters induced by the 
minimal torsion coupling to fermions. We shall review Holst gravity and obtain several coefficients from the Lagrangian. 
A similar expression will be obtained for torsion trace $T$ and its respective mass as $M_{T}$. Holst term is given by: 
${\cal{H}}={\epsilon}^{ijkl}R_{ijkl}$. Throughout this paper, we use $i,k,l=0,1,2,3$ and $R_{ijkl}$ are the components 
of torsion. Throughout this paper, we use $i,j,k,l=0,1,2,3$ and $R_{ijkl}$ are the components of torsion. Recently, 
Karananas et al.\,\cite{13} formulated a purely gravitational origin for axions in minimal EC, incorporating affine 
connection with torsion trace. In this paper, using Holst gravity, we derive simpler Lagrangians bounded by LHC hadron 
collider experiment proposals for torsion detection by Almeida et al.\,\cite{3}. For this purpose, we assume the 
Riemannian metric $g_{ij}\approx{{\eta}_{ij}}$, where ${\eta}_{ij}$ is the Minkowski 4-dimensional flat metric. 

The reminder of thus paper is organized as follows. Section 2 describes a technique used to derive field Lagrangians and 
coefficients from field equations, which has been employed to study dark matter from Higgs inflation. By extending the 
Einstein-Cartan-Holst action to the Einstein-Cartan-Holst-Zanelli action, this approach addresses the coupling between 
the $Z$ boson mass spectrum and torsion, offering potential insights for experimental detection at CERN-LHC. Section 3 
explores the impact of axion transmutation into torsion in Holst gravity, specifically examining how torsion and axion 
masses influence magnetic helicity instability, and also derives equations for dark photon interactions, highlighting the 
role of torsion mass and BI parameter in axion oscillations and dark photon frequency. Section 4 is left for summary and 
outlook.

\section{\textbf{Einstein-Cartan-Holst Lagrangian for dark photons}}
The purpose of this section is to present a simple technique used by several authors to obtain field Lagrangians 
and determine coefficients from the corresponding field equations. This technique has been employed by Rigouzzo and 
Zell\,\cite{14} to obtain dark matter from Higgs inflation. By extending the ECH action to 
Einstein-Cartan-Holst-Zanelli\,(ECHZ) action, it addresses problems such as finding the coupling constant between the 
$Z$ boson's mass spectrum and torsion, potentially offering a way to detect observable parameters at CERN-LHC. Here, 
we assume a minimal coupling between the $Z$ boson and torsion via the Holst term.
Let us consider the ECHZ action given by
\begin{equation}
{\cal{L}}_{ECHZ}= {\cal{L}}_{EC}+{\cal{L}}_{H}+{\cal{L}}_{NY},
\label{3}
\end{equation}
where $\cal{L}_{ECHZ}$ is composed of the Einstein-Cartan Lagrangian $\cal{L}_{EC}$, the Holst term $\cal{L}_{H}$, 
and the Nieh-Yan term $\cal{L}_{NY}$, which describes specific topological effects in the theory, often relating to the 
coupling of torsion and curvature in ways that are topologically invariant. Within this general framework, the Lagrangian 
for the dark photon ${\gamma}'$ can be expressed as:
\begin{equation}
{\cal{L}}_{{\gamma}'}\supset{\frac{1}{2}({a_{0}}F^{2}+a_{2}aF{\tilde{F}}+\frac{a_{3}}{2}({D}a)^{2}+a_{4}e^{2}A^{2})},
\label{4}
\end{equation}
where $\tilde{F}$ is the dual of the electromagnetic form $F$. 
When we substitute the expression for covariant derivative $D$, which is 
\begin{equation}
D= \partial-ieA-ig\gamma_{5}-ifT ,
\label{5}
\end{equation}
where $\partial$ is the standard partial derivative representing differentiation in flat spacetim, 
$-ieA$ presents the interaction with the electromagnetic potential $A$, $-ig{\gamma}_{5}$ accounts for 
interaction with an axial vector field, $-ifT$ represents the interaction with the torsion field, $g$ and 
$f$ are the coupling constants, then we get
\begin{equation}
{\cal{L}}_{a{\gamma}'}\sim{-\frac{1}{2}{a_{1}}F^{2}+a_{2}aF{\tilde{F}}+[\frac{a_{3}}{2}+a_{5}g^{2}+ia_{1}g{\gamma}^{5}]
({\partial}a)^{2}-\frac{1}{2}m_{a}^{2}a^{2}} .
\label{6}
\end{equation}
where the first term represents the kinetic energy of the electromagnetic field $F$, describing its propagation and interactions 
in space; the second term illustrates the interaction between the scalar field $a$, the electromagnetic field, and its dual, 
introducing effects that can break certain symmetries, the third term describes the kinetic energy of the scalar field, 
including modifications due to interactions with other fields, thereby explaining the field's dynamics, and the fourth term 
is the mass term for the scalar field, crucial for determining its mass properties and behavior.
Now, considering the torsion Lagrangian:
\begin{equation}
{{\cal{L}}_{T}=[\frac{1}{\beta}(a_{7}+ia_{11}fg)T+a_{10}egA{\gamma}^{5}]{\partial}{a}+a_{4}e^{2}A^{2}}+a_{6}f^{2}{T}^{2} .
\label{7}
\end{equation}
The axial torsion $S$ is given, following Duncan et al.\,\cite{16} by ${{\partial}a}$ as a transmutation rule between axion 
and torsion. The variation of the torsion trace $T$ of the sum of the Lagrangians is
\begin{equation}
{\cal{L}}={\cal{L}}_{T}+{\cal{L}}_{a{\gamma}'} .
\label{8}
\end{equation}
The relation between the torsion trace and axion\,\cite{15} is not given a priori, but derived from the field equations. 
From the variation of the Lagrangian, we derive the relationship between the torsion trace $T$ and the axion field $a$,
\begin{equation}
T=-\frac{1}{\beta}[\frac{a_{7}+ia_{11}fg}{a_{6}f^{2}}]{\partial}a .
\label{9}
\end{equation}
This equation shows that the torsion trace is proportional to the derivative of the axion field, with the proportionality 
factor involving the coupling constants $\beta, a_{7}, a_{11}, f$ and $a_{6}$. Assuming all coefficients $a_{A}$ (where 
$A=1,...,11)$ are real, the BI parameter is real, and if $f$ and $g$ are real numbers, we assume that $a_{11}$ vanishes. 
Therefore, Equation\,(\ref{9}) simplifies to
\begin{equation}
T=-\frac{1}{\beta}\frac{a_{7}}{a_{6}f^{2}}{\partial}a  .
\label{10}
\end{equation}
This equation can be rewritten in terms of the oscillatory term as
\begin{equation}
{\partial}T=-\frac{1}{\beta}\frac{a_{7}}{a_{6}f^{2}}{\Box}a .
\label{11}
\end{equation}
The step from $\partial({\partial}a)$ to ${\Box}a$ assumes that we are working within the context of field 
equations where the d'Alembertian operator is a common notation. The expression\,(\ref{11}) will be substituted into the axion 
equation to derive an expression dependent on the Immirzi parameter but independent of $T$. By varying $\delta{a}$, one 
gets 
\begin{equation}
[1+M_{S}^{2}+a_{5}g^{2}+ia_{1}g{\gamma}^{5}-\frac{a_{7}}{2a_{6}{\beta}^{2}f^{2}}]\Box{a}+m_{a}^{2}a=-a_{2}F{\tilde{F}} ,
\label{12}
\end{equation}
where $a_{3}$ is taken as the square of the axial torsion pseudo-vecto $S$. The term $ia_{1}g{\gamma}^{5}$ includes an 
imaginary unit $i$, the coupling constant $a_{1}$, another coupling constant $g$, and $\gamma^{5}$, which is related to 
the axial vector field interaction, and $-\frac{a_{7}}{2a_{6}{\beta}^{2}f^{2}}$ represents a correction to the interaction 
term.

Next, by varying the electromagnetic\,(EM) potential $A$, we obtain
\begin{equation}
{\partial}F-a_{2}S{\tilde{F}}-2a_{4}e^{2}A-a_{10}eg{\gamma}_{5}S= J^{5} ,
\label{13}
\end{equation}
where we consider the torsion coupling constant $a_{10}$ to be zero to simplify the equation. The chiral current $J_{5}$ 
arises from the term  $J_{5}A$ in the interaction Lagrangian. From Equation (13), if we do not want torsion to couple 
directly with electromagnetic fields or only in a fermionic current, we must choose $a_{10}=0$. However, this forces $a_{2}$ 
to vanish as well.

Now, let's determine the coefficients from the above equations. Starting with the EM equations, we 
notice that the coefficient
\begin{equation}
a_{4}=-\frac{m^{2}_{{\gamma}'}}{e^{2}}\label{14} .
\end{equation}
This expression is interesting as it is analogous to the expressions for torsion-fermion couplings presented by 
arman et al.\,\cite{1}. From the total Lagrangian, we can show that
\begin{equation}
a_{2}= \frac{{g_{a\gamma'}}}{f_{a}} ,
\label{15}
\end{equation}
where $f_{a}$ is the axion decaying constant into photons. Additionally, we obtain the following coefficients
\begin{equation}
a_{6}=\frac{M^{2}_{T}}{f^{2}} ,
\label{16}
\end{equation}
\begin{equation}
a_{5}=\frac{M^{2}_{S}}{g^{2}} .
\label{17}
\end{equation}
Another important coefficient can be determined from the equation for torsion oscillations, which is significant in dark 
torsion in bouncing cosmologies\,\cite{5}. This coefficient is
\begin{equation}
a_{7}=\frac{1(\mathrm{TeV})}{{\beta}{M}_{T}^{2}} ,
\label{18}
\end{equation}
allowing us to to estimate the loop quantum gravity-inspired BI parameter $\beta$ provided we have 
$a_{7}$. Inverting this equation yields
\begin{equation}
{M^{2}_{S}}=\frac{1}{\sqrt{2}{\beta}M_{T}^{2}} .
\label{19}
\end{equation}
Now notice that Equatioin\,(19) can still be expressed as 
\begin{equation}
{(M_{S}M_{T})}^{2}=\frac{1}{\sqrt{2}{\beta}} ,
\label{20}
\end{equation}
where we take $a_{7}=1$. From this generalized ECH Lagrangian, assuming the average 
torsion trace and axial masses are approximately equal, we may expressed a final interesting formula case as
\begin{equation}
{M}^{4}=\frac{1}{2{\beta}^{2}} ,
\label{21} 
\end{equation}
which reduces to
\begin{equation}
{M}^{2}\approx {\frac{1}{\sqrt{2}\sqrt{2}{\beta}}} .
\label{22} 
\end{equation}
Here, $M$ denotes the average mass, either for torsion trace or axial torsion masses. Using this expression, 
we can calculate the torsion mass from the BI parameter and vice-versa.For example, consider the case by Panza et al.\,\cite{6}, with an upper bound of the BI parameter equal to 1.185. Substituting this into Equation (21) yields 
 $M=0.775\,\mathrm{TeV}=775\,\mathrm{GeV}$. Compared to the Higgs boson mass of around 125 GeV, this shows that the 
massive torsion is much heavier than the Higgs boson, yet 0.775 TeV is still considered light torsion. Using the result 
obtained by the author, we estimate 0.285 for the BI parameter, implying an average torsion mass of 1.51 TeV. Although 
heavier than the torsion produced by Panza et al.\,\cite{6} at TeV scales with a quadratic Lagrangian in the Riemann-Cartan 
curvature, it is still considered light torsion. It is noteworthy that at the EC limit, when the BI parameter 
approaches infinity, there is no torsion mass in the original form of EC theory by Cartan, Trautman, Sciama, and Kibble. 
This constant is better understood by considering analogous expressions obtained by Barman et al.\,\cite{1}, taking into 
account the interaction Lagrangian:
\begin{equation}
{\cal{L}}_{int}=-\frac{{\eta}^{2}}{M^{2}_{S}}(\bar{f}{\gamma}^{i}{\gamma}^{5}f)^{2} ,
\label{20}
\end{equation}
 yielding
\begin{equation}
\frac{{M}_{s}}{\eta}\ge{1.7\,\mathrm{TeV}} ,
\label{21}
\end{equation}
where ${\eta}$ is the torsion fermion coupling constant given by 
\begin{equation}
{\cal{S}}= \int{d^{4}x{(\bar{f}{\gamma}^{i}D_{i}f-mf\tilde{f})}} .
\label{22}
\end{equation}
We also note that the four-dimensional curl $S_{ij}$ is expressed as
\begin{equation}
S_{ij}=2{\partial}_{[i}S_{j]} ,
\label{23}
\end{equation}
which vanishes when the axial torsion vector is identified with the gradient of the axion field. This is interesting 
because, according to Carroll et al.\,\cite{15}, this term cannot be present along the $S^{2}$ simultaneously in the 
Lagrangian density due to renormalization issues. The main results are summarized in the following table.
\begin{table}[h!]
\centering
\begin{tabular}{ccc}
\hline
               Einstein-Cartan   & $M$(TeV) &   BI(LQG)\\
\hline
Panza et al (2016)&  0.775     &1.185 \\
de Andrade (2024)&   1.51        &0.285 \\  
\hline
\end{tabular} \caption{Barbero-Immirzi (BI) parameter from light torsion masses ($M$) in distinct physical backgrounds}
\label{table:formulas}
\end{table}
The Barbero-Immirzi parameter can be computed from Equation\,(\ref{18}) by taking the $a_{7}={1\,\mathrm{TeV}}^{2}$. This results in $\beta=0.285$, well within the range obtained by Panza et al.\,\cite{6} using a quadratic curvature Lagrangian at TeV 
scales, where $0\le{\beta}\le{1.185}$. From the well-known value of the BI parameter for black hole entropy in quantum 
gravity, $0.273$, our formula yields a torsion mass of $1.71$\,TeV.

\section{Dark photons and magnetic helicity instabilities in Holst gravity}
Recently, Agrawal et al.\,\cite{12} investigated the abundance of tachyonic instability and the production of dark photon 
dark matter in the universe as the axion oscillates. In this section, we consider the relation of axion transmutation into 
torsion in Holst gravity, as discussed by Duncan et al.\,\cite{16}, to study the influence of torsion mass and axion mass 
on magnetic helicity instability. We start with the interaction Lagrangian
\begin{equation}
{\cal{L}}_{int}\supset{\frac{\phi}{4f_{a}}F_{D}{\tilde{F}}_{D}} ,
\label{27}
\end{equation}
where $f_{a}$ is the decay coupling of the axion to photons, and and the subscript $D$ on the electromagnetic tensor 
$F$ indicates that we are talking about dark photons. To derive the dark photon equation with torsion masses in Holst 
gravity with non-minimal coupling, we write the Lagrangian with the Holst term and dark massive photon term as
\begin{equation}
{\cal{L}}_{Holst-{\gamma}'}\supset{a({\partial}{\phi})^{2}+bTS+cT^{2}+dS^{2}+e{\phi}F_{D}{\tilde{F}}_{D}+fA^{2}-V(\phi)-
\frac{1}{4}F^{2}} ,
\label{28}
\end{equation}
where $a({\partial}{\phi})^{2}$ represents the kinetic energy of the axion field $\phi$, $bTS$ is the mixed Holst term, 
indicating the coupling between the torsion trace $T$ and the axial torsion $S$, $cT^{2}$ describes the self-interaction of 
the torsion trace, representing the energy density of the torsion trace, $e{\phi}F_{D}{\tilde{F}}_{D}$ describes the 
self-interaction of the axial torsion, representing the energy density of the axial torsion, $e{\phi}F_{D}{\tilde{F}}_{D}$ 
represents the interaction between the axion field and the dark photon field\footnote{Here, $F_{D}$ is the electromagnetic 
tensor of the dark photon, and ${\tilde{F}}_{D}$ is its dual tensor. This term indicates the coupling between the axion 
field and the dark photon field.}, $fA^{2}$ represents the mass term for the dark photon field, indicating the mass 
energy of the dark photon field, $-V(\phi)$ is the potential energy term for the axion field, and $\frac{1}{4}F^{2}$ 
represents the kinetic energy of the standard electromagnetic field. The coefficients are defined as 
\begin{equation}
a= \frac{1}{2};  \newline
b=-\frac{m^{2}_{P}}{2\beta};
c=\frac{m^{2}_{T}}{2}; d= \frac{m^{2}_{S}}{2}; e=\frac{1}{4f_{\phi}} ,
\label{29}
\end{equation}
whereas the potential $V(\phi)$ is approximated by 
\begin{equation}
V(\phi)= \frac{m^{2}_{\phi}}{2} .
\label{30}
\end{equation}
Of course we also use Duncan et al.'s assumption in QED that $S= {\partial}{\phi}$\,\cite{16}. By analogy what we did in the 
last section we perform the variation of the last Lagrangian to $T$ trace torsion vector to check the relationship between 
the trace torsion vector and axion $\phi$. This results in
\begin{equation}
T=-\frac{b}{c}{\partial}{\phi} .
\label{31}
\end{equation}
 Equation\,(\ref{31}) shows that the torsion trace four-vector is given by the gradient of the axion, similar to 
Duncan et al.\,\cite{16}, but now the coefficient is proportional to $\frac{m^{2}_{P}}{m^{2}_{T}}$. It is important to note 
that the vector $T$ was not present in Duncan et al.\,\cite{16}. Substituting Equation\,(\ref{31}) into the Lagrangian, 
the mixed Holst term $TS$ is replaced by a term proportional to the axion gradient. The presence of torsion in the 
field equations is represented solely by their mass spectrum as
\begin{equation}
{\cal{L}}_{{\gamma}'{\phi}A}\supset{g({\partial}\phi)^{2}+e{\phi}F_{D}{\tilde{F}}_{D}+fA^{2}_{D}-\frac{m^{2}_{\phi}}{2}
{\phi}^{2}-\frac{1}{4}F^{2}} ,
\label{32}
\end{equation}
where $g$ constant now is  expressed in terms of other constants by
\begin{equation}
g= [a(1+\frac{1}{c})+d-\frac{1}{2}\frac{b^{2}}{c}] .
\label{33}
\end{equation}
The method used by Dombriz et al.\,\cite{10} to compute the coefficient $g$ of the final Lagrangian in terms of 
physical parameters can be applied here
\begin{equation}
g\approx{\frac{1}{2}[1-m^{2}_{S}+\frac{m^{4}_{P}}{2{\beta}^{2}m^{2}_{T}}]} .
\label{34}
\end{equation}
This expression can be further approximated to
\begin{equation}
g\approx{\frac{1}{2}[-m^{2}_{S}+\frac{m^{4}_{P}}{2{\beta}^{2}m^{2}_{T}}]}  .
\label{35}
\end{equation}
From Equation\,(\ref{34}), the BI parameter of loop quantum gravity can be obtained by setting the sum of the second and 
third terms in $g$ to zero. This will reduce the coefficient $g$ to 1, as in the regular Klein-Gordon equation. The BI 
parameter then becomes
\begin{equation}
\sqrt{2\beta}= \frac{m^{2}_{P}}{m^{2}_{T}} ,
\label{36}
\end{equation}
where we have used the approximation $m_{T}\approx{m_{S}}$. Using the LHC table of axial torsion by Almeida et al.\,\cite{3}, 
we can create a histogram (Figure 1) for the BI parameter and the trace torsion mass in GeV.
\begin{figure}[h]
    \centering
    \includegraphics[width=0.9\linewidth]{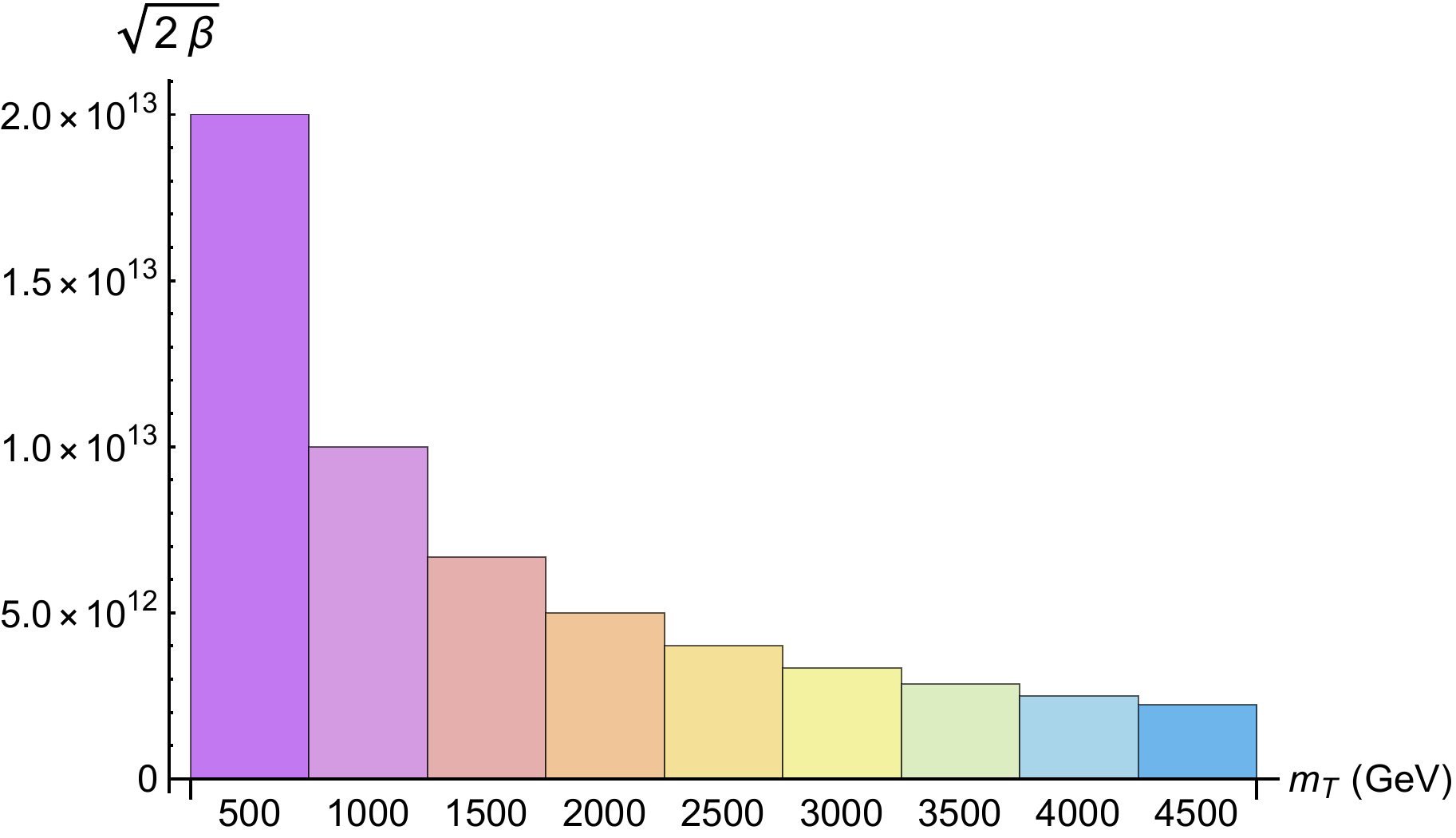}
    \caption{The relation between the true BI parameter and the trace torsion mass. Here $\gamma={\beta}^{-1}$.}
    \label{fig:enter-label}
\end{figure}

Now let us built the both equations for coupling system of dark photon and axions which induce axion oscillations from 
tachyonic instability. By varying the Lagrangian with respect to $\phi$, one obtains the axion equation
\begin{equation}
{\Box}\phi+\frac{m^{2}_{\phi}}{g}{\phi}= \frac{e\textbf{E}\cdot\textbf{B}}{g} .
\label{37}
\end{equation}
Here, ${\Box}\phi$ represents the d'Alembertian operator applied to the axion field $\phi$, $\frac{m^{2}_{\phi}}{g}{\phi}$ 
is the mass term, and $\frac{e \textbf{E} \cdot \textbf{B}}{g}$ represents the interaction term with the electromagnetic fields.

In the chirality-free case where the chiral term $\textbf{E}\cdot\textbf{B}$ vanishes, we can simplify and obtain
\begin{equation}
{\omega}_{\phi}= \frac{m_{\phi}}{\sqrt{g}}.
\label{38}
\end{equation}
Here, ${\omega}_{\phi}$ is the oscillation frequency of the axion. Since $g$ can be expressed in terms of the BI parameter 
and torsion trace masses, the axion oscillation depends on the BI parameter and torsion mass. To obtain this solution, we 
use the following ansatz
\begin{equation}
{\phi}= {\phi}_{0}\exp{[i{\omega}_{\phi}t]} .
\label{39}
\end{equation}
This ansatz characterizes the oscillation of the axion. Next, we find the second part of the solution by approximating 
the equation as
\begin{equation}
\phi= \frac{\textbf{E}\cdot\textbf{B}}{4m^{2}_{\phi}f_{\phi}} .
\label{40}
\end{equation}
From this expression, it is clear that to obtain the second axion solution, we need to explicitly determine the vector 
potential $A$, since the electric and magnetic vector fields can be obtained as
\begin{equation}
\textbf{E}= -{\nabla}{\psi}-{\partial}_{t}\textbf{A} ,
\label{41}
\end{equation}
and 
\begin{equation}
\textbf{B}= {\nabla}\times{\textbf{A}} .
\label{42}
\end{equation}
Here, $\psi$ is the electric scalar potential and should not be confused with the axion field. The helicity parameter 
$\lambda$ shall be given by the equation ${\nabla}\times{\textbf{A}}={\lambda}\textbf{A}$. Therefore by substitution of 
these expressions into the right-hand side of Equation\,(\ref{39}) one gets
\begin{equation}
\phi =-\frac{1}{4f_{\phi}m^{2}_{\phi}}{[h+ \frac{d}{dt}\textbf{A}^{2}]} ,
\label{43}
\end{equation}
where $h ={\textbf{E}}_{0}\cdot\textbf{A}$ is the magnetic helicity, and ${\textbf{E}}_{0}=-{\nabla}\psi$ is the 
electrostatic field. Since the chirality term is written in terms of the magnetic potential, to solve the magnetic 
wave field, we use the following equation
\begin{equation}
{\partial}_{i}F^{ij}+({\partial}_{i}\phi){\tilde{F}}^{ij}= m^{2}_{{\gamma}'}A^{i} .
\label{44}
\end{equation}
From this Maxwell-like equation one obtains
\begin{equation}
\ddot{\textbf{A}}+ [k^{2}+m^{2}_{{\gamma}'}-\frac{\dot{\phi}{\lambda}}{f_{\phi}}]\textbf{A}=0 .
\label{45}
\end{equation}
A simple solution can be obtained by using the approximation ${\cal{O}}(A^{3})\approx{0}$. With this approximation the 
time derivative of the axion scalar reduces to 
\begin{equation}
\dot{\phi}=-\frac{\dot{h}}{4f_{\phi}m^{2}_{\phi}} .
\label{46}
\end{equation}
Substitution of this expression into the magnetic wave equation yields
\begin{equation}
\ddot{\textbf{A}}+[k^{2}+{m^{2}_{\phi}}-\frac{{\lambda}\dot{h}}{16f^{2}_{\phi}m^{4}_{\phi}}]\textbf{A}=0 .
\label{47}
\end{equation}
From this expression we immediatly derive the dark photon frequency as
\begin{equation}
{\omega}^{2}_{D}=[k^{2}+m^{2}_{{\gamma}'}-\frac{{\lambda}\dot{h}}{16f^{2}_{\phi}m^{4}_{\phi}}] .
\label{48}
\end{equation}
From this expression one may notice that the dark photon frequency is severely damped if both helicities, $h$ and $\lambda$ 
have the same sign, and is enhanced in the other case. This relationship highlights the intricate interplay between the axion 
field and electromagnetic fields in influencing the properties of dark photons. Understanding this behavior is crucial for 
predicting observable effects in experiments aimed at detecting dark matter candidates, such as axions and dark photons. 
Future research may explore how variations in the BI parameter and torsion mass could provide deeper insights into the 
fundamental interactions within Holst gravity and their implications for cosmology and particle physics.
\section{Summary and Outlook}
In this paper, we introduced a new probe for the BI parameter in the context of Holst gravity with a 
portal to dark photons. Frequencies of axion and dark photons reveal that in Holst gravity, axions induced by torsion may 
be produced by the damping of dark photons. We notice that in our example, there seems to be a decoupling between torsion 
and axions, as torsion does not explicitly appear in the field equations, but this was masked by the constraints on the BI 
parameter, which appears in terms of Planck and torsion masses. Our findings reveal that if we do not assume the constraint 
chosen for the BI parameter, torsion masses would appear in the dark photon field equations. Moreover, the relationship 
between axion and torsion suggests a correlation between their masses, similar to the relationship between dark photon and 
axion masses, thereby implying a link between dark photon and torsion masses. This relationship has also been investigated 
by Gao et al.\,\cite{17} in a recent study. This study may broaden the understanding of Holst gravity, as well as provide 
crucial insights into the interplay between torsion, dark photons, and axions in the cosmological context.

Future research should further explore these interactions to enhance our understanding of the complex behaviors within 
Holst gravity and their implications for cosmology and particle physics. Specifically, investigating the coupling between 
the $Z$ boson mass spectrum and torsion within the extended Einstein-Cartan-Holst-Zanelli action could offer potential 
insights for experimental detection at CERN-LHC. Additionally, examining how torsion and axion masses influence magnetic 
helicity instability and deriving equations for dark photon interactions could provide valuable information on the role of 
torsion mass and BI parameter in axion oscillations and dark photon frequency.
\section{Acknowledgments}
I thank S. Zell, M. Shaposhnikov and A. Nepomuceno for helpful discussions on the subject of this paper. Financial support 
from Universidade do Estadoof Rio de Janeiro (UERJ) is grateful acknowledged. This work was performed under the auspices of 
Major Science and Technology Program of Xinjiang Uygur Autonomous Region through No.2022A03013-1.

\end{document}